\documentclass[
reprint,amsmath,amssymb,aps,prdl
]{revtex4-1}

\usepackage{epsfig}
\usepackage{dcolumn}

\usepackage{lineno}
\usepackage{bm}
\usepackage{physics}
\usepackage{color}
\usepackage{graphics}
\graphicspath{{graphs/}}
\usepackage{subfigure}
\usepackage{overpic}  % 
\usepackage{times} %
\usepackage[english]{babel}
\usepackage{amsmath}
\usepackage{amssymb}
\usepackage{mathrsfs} % 
\usepackage{breqn}  % 
\usepackage[colorlinks,linkcolor=blue,citecolor=blue]{hyperref}
\usepackage{lineno}
\usepackage{multirow}
\usepackage{float}
\usepackage{hyperref}
\newcommand{\BESIIIorcid}[1]{\href{https://orcid.org/#1}{\hspace*{0.1em}\raisebox{-0.45ex}{\includegraphics[width=1em]{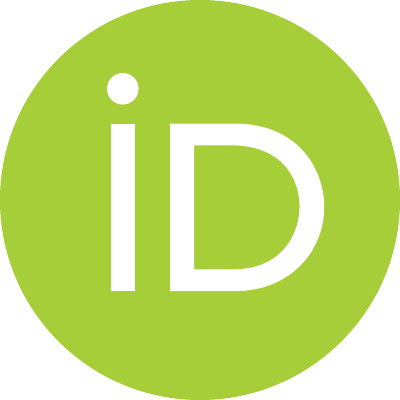}}}}
\hypersetup{
    colorlinks=true,
    linkcolor=blue,
    filecolor=blue,      
    urlcolor=blue,
    citecolor=blue,
}

\begin{document}

\title{Measurement of Born cross section for $e^+e^-\to K_S^0\bar{\Xi}^+\Sigma^-$ at $\sqrt{s} = 3.51-4.95$ GeV
}
\author{BESIII Collaboration}
\thanks{Full author list given at the end of the Letter}

\affiliation{}

\begin{abstract}
Using $e^+e^-$ collision data collected with the BESIII detector at the BEPCII collider corresponding to a total integrated luminosity of 44~fb$^{-1}$,
we present the first measurement of the Born cross sections for the process $e^+e^-\to K_S^0\bar{\Xi}^+\Sigma^-$ at 56 center-of-mass energies from 3.510 to 4.951~GeV. By fitting the dressed cross sections of $e^+e^-\to K_S^0\bar{\Xi}^+\Sigma^-$ with the assumption of a power-law function plus a charmonium(-like) resonance, i.e. $\psi(3770)$, $\psi(4040)$, $\psi(4160)$, $Y(4230)$, $Y(4360)$, $\psi(4415)$, $Y$(4500), $Y(4660)$, and $Y$(4710), no significant signal of any charmonium(-like) state decaying into the $K_S^0\bar{\Xi}^+\Sigma^-$ is observed.
Upper limits on the product of the electronic width and branching fraction at the 90\% confidence level are given for each resonance. Combining this result with the previous measurement of the isospin-symmetric process $e^+e^-\to K^{-} \bar{\Xi}^{+} \Sigma^{0} $, the ratio of the Born cross sections,
$R=\sigma^{B}(e^+e^-\to K_S^0\bar{\Xi}^+\Sigma^-)/$$\sigma^{B}(e^+e^-\to K^-\bar{\Xi}^+\Sigma^0)$, is found to be consistent with unity.

\end{abstract}

\maketitle
%\linenumbers 
Studies of baryonic final states in $e^+e^-$ annihilation provide essential input for understanding the production mechanisms of charmonium(-like) states.
In particular, investigating the production of charmonium(-like) states above the open-charm threshold and their hadronic decays offers a sensitive test of QCD calculations and a unique probe of XYZ states~\cite{Brambilla:2010cs, Briceno:2015rlt,Wang:2025dur,Bai:2026atm}.
Potential models predict five vector charmonium states between 3.773 and 4.7~GeV/$c^2$: the 3$S$, 2$D$, 4$S$, 3$D$, and 5$S$ states~\cite{Brambilla:2010cs}.
However, experiments have observed more vector states in this region than predicted.
Several states, such as the $\psi(4040)$, $\psi(4160)$, and $\psi(4415)$, are seen in hadronic cross sections~\cite{ParticleDataGroup:2022pth}.
In addition, via initial-state radiation (ISR), the \textsc{BaBar} and Belle collaborations have observed three charmonium-like states: $Y(4230)$, $Y(4360)$, and $Y(4660)$~\cite{BaBar:2005hhc,BaBar:2006ait,Belle:2007umv,Belle:2007dxy,Belle:2008xmh,BaBar:2012vyb,Belle:2013yex,BaBar:2012hpr,Belle:2014wyt}.
Direct production at the CLEO~\cite{CLEO:2006ike} and BESIII~\cite{BESIII:2014rja,BESIII:2024ths} experiments has confirmed the $Y(4230)$, $Y(4360)$, and $Y(4660)$ states;
these cannot be classified within the simple $c\bar{c}$ quark model, and their nature remains unresolved. 
Refs.~\cite{Wang:2019mhs,Qian:2021neg} suggest that these states are considered pure charmonium states, provided that their baryonic decays can be observed. This situation highlights the incomplete understanding of the strong interaction, particularly in its nonperturbative aspects.
Although BESIII has performed extensive studies in this energy region~\cite{BESIII:2013ujm,BESIII:2017kqg,BESIII:2019cuv,BESIII:2021aer,BESIII:2021cvv,BESIII:2021ccp,BESIII:2023rse,BESIII:2023rwv,BESIII:2024umc,BESIII:2024ogz,BESIII:2024ues,BESIII:2024gql,BESIII:2025lbj,Zhang:2025qmo,BESIII:2026hie,Zhang:2026qjt,BESIII:2026oyx,BESIII:2026ont,BESIII:2026lcb}, only $\psi(3770)\to\Lambda\bar{\Lambda}$~\cite{BESIII:2021ccp} and $\psi(3770)\to\Xi^-\bar{\Xi}^+$~\cite{BESIII:2023rse} have been reported.
To clarify the nature of these charmonium(-like) states, precise measurements of exclusive $e^+e^-\to$~baryonic cross sections above the open-charm threshold are essential.

This Letter reports the first measurement of the Born cross sections for $e^+e^-\to K_S^0\bar{\Xi}^+\Sigma^-$ (hereafter the charge-conjugate model is implied by default).
The ratio of the Born cross sections to those for $e^+e^-\to K^-\bar{\Xi}^+\Sigma^0$ is also determined.
The data correspond to an integrated luminosity of 44~fb$^{-1}$ of $e^+e^-$ collisions at center-of-mass (c.m.) energies $\sqrt{s}$ between 3.510 and 4.951~GeV~\cite{Ablikim:2013ntc, BESIII:2015qfd, BESIII:2022dxl, BESIII:2022ulv, BESIII:2024lks, BESIII:2024lbn}, collected by the BESIII detector~\cite{BESIII:2009fln} at the BEPCII collider~\cite{Yu:2016cof}.

Event candidates for $e^+e^-\to K_S^0\bar{\Xi}^+\Sigma^-$ are selected via a partial-reconstruction technique.
Only $K_S^0$ and $\bar\Xi^+$ are reconstructed, through the decays $K_S^0 \to \pi^+\pi^-$ and $\bar{\Xi}^+ \to \bar{\Lambda}\pi^+$ with $\bar{\Lambda} \to \bar{p}\pi^+$; the $\Sigma^-$ is inferred from the recoil against the reconstructed $K_{S}^0\bar\Xi^{+}$ system.
Monte Carlo (MC) samples are produced with a {\sc geant4}-based~\cite{GEANT4:2002zbu} package that includes the BESIII geometry and a realistic detector response; these are used to determine efficiencies and to estimate background contributions.
Detection efficiencies for $e^+e^-\to K_S^0\bar{\Xi}^+\Sigma^-$ are obtained from $4\times10^5$ phase-space (PHSP) events generated at each energy point with the $\textsc{kkmc}$ generator~\cite{Jadach:1999vf, Jadach:2000ir}, which incorporates beam-energy spread and ISR corrections.
Subsequent decays are simulated with $\textsc{evtgen}$~\cite{Ping:2008zz, Lange:2001uf} using branching fractions from the Particle Data Group~(PDG)~\cite{ParticleDataGroup:2022pth}; unmeasured modes are handled by $\textsc{lundcharm}$~\cite{Chen:2000tv}.
An inclusive MC sample generated at $\sqrt{s} = 3.773$~GeV is used to estimate backgrounds; it includes $D\bar{D}$ production (with quantum coherence for neutral $D$ mesons), non-$D\bar{D}$ decays of $\psi(3770)$, ISR production of $J/\psi$ and $\psi(3686)$, and continuum processes incorporated in {\sc kkmc}~\cite{Jadach:1999vf, Jadach:2000ir}.

Charged tracks are reconstructed in the multilayer drift chamber (MDC) within the angular region $\lvert\cos\theta\rvert <0.93$, where $\theta$ is the polar angle with respect to the symmetry axis of the MDC. %$z$ axis in the laboratory system.
At least three positive-charged tracks and two negative-charged tracks are required to be reconstructed in the MDC.
Particle identification~(PID) for charged tracks combines measurements of the specific ionization energy loss in the MDC~(d$E$/d$x$) and the flight time in the time-of-flight system 
to form likelihoods $\mathcal{L}(h)~(h=p,K,\pi)$ for each hadron $h$ hypothesis.
Tracks are identified as protons when the proton hypothesis has the greatest likelihood ($\mathcal{L}(p)>\mathcal{L}(K)$ and $\mathcal{L}(p)>\mathcal{L}(\pi)$), and as pions when $\mathcal{L}(\pi)>\mathcal{L}(p)$ and $\mathcal{L}(\pi)>\mathcal{L}(K)$.
Events with at least one $\bar{p}$, three $\pi^+$s, and one $\pi^-$ are kept for further analyses.

To reconstruct the $K_S^0$ candidate, a vertex fit and secondary vertex fit are applied to all $\pi^+\pi^-$ combinations with a mass difference minimization strategy, $\delta_{K_S^0}=\lvert M_{\pi^+\pi^-}-m_{K_S^0}\rvert$, which reconstructs the displaced decay vertex and provides the decay length used in the subsequent selection. Here, $M_{\pi^+\pi^-}$ denoted the invariant mass of the $\pi^+\pi^-$ combination, and $m_{K_S^0}$ is the nominal $K^0_S$ mass taken from the PDG~\cite{ParticleDataGroup:2022pth}. To suppress the potential background candidates, the $\pi^+\pi^-$ invariant mass is required to be within 12~MeV/$c^2$ of the nominal $K_S^0$ mass, and the $K_S^0$ decay length over its uncertainty is required to be $L_{K_S^0}$/$\Delta L_{K_S^0} > 2$~\cite{BESIII:2021haw}. 
To reconstruct the $\bar{\Lambda}$ candidates, a similar secondary vertex fit is applied to all $\bar{p}\pi^+$ combinations. 
The invariant mass of $\bar{p}\pi^+$ is required to be $\lvert M_{\bar{p}\pi^+}-m_{\bar{\Lambda}}\rvert < 5~ \rm{MeV} /\textit{c}^2$~\cite{BESIII:2021ccp}, and the decay length of $\bar{\Lambda}$ is required to be larger than zero~\cite{BESIII:2024umc}. Here, $M_{\bar{p}\pi^+}$ denoted the invariant mass of the $\bar{p}\pi^+$ combination, and $m_{\bar{\Lambda}}$ is the nominal $\bar{\Lambda}$ mass taken from the PDG~\cite{ParticleDataGroup:2022pth}.

The $\bar{\Xi}^+$ candidates are formed by combining an additional $\pi^+$ with a reconstructed $\bar{\Lambda}$ and performing a secondary-vertex fit.
The $\bar{\Lambda}\pi^+$ invariant mass must satisfy $| M_{\bar{\Lambda}\pi^+}-m_{\bar{\Xi}^+}|<8$~MeV/$c^2$, and the decay length is required to be positive.
The best $\bar{\Lambda}$ and $\bar{\Xi}^+$ candidates are selected by minimizing $\delta_{\rm min}= \sqrt{| M_{\bar{p}\pi^+}-m_{\bar{\Lambda}}|^2 + | M_{\bar{\Lambda}\pi^+}-m_{\bar{\Xi}^+}|^2}$. Here, $M_{\bar{\Lambda}\pi^+}$ denoted the invariant mass of the $\bar{\Lambda}\pi^+$ combination, and $m_{\bar{\Xi}^+}$ is the nominal $\bar{\Xi}^+$ mass taken from the PDG~\cite{ParticleDataGroup:2022pth}.
The $\Sigma^-$ is inferred from the recoil mass against the $K_S^0\bar{\Xi}^+$ system:
\begin{equation}
    M_{K_S^0\bar{\Xi}^+}^{\rm recoil} = \sqrt{(\sqrt{s}-E_{K_S^0\bar{\Xi}^+})^2-| {\vec{p}_{K_S^0\bar{\Xi}^+}|^2}},
\end{equation}
where $E_{K_S^0\bar{\Xi}^+}$ and $\vec{p}_{K_S^0\bar{\Xi}^+}$ are the energy and momentum of the selected $K_S^0\bar{\Xi}^+$ system in the $e^+e^-$ c.m.\ frame, respectively.
After applying all selection criteria to the inclusive MC sample at $\sqrt{s}=3.686$ and $\sqrt{s} = 3.773$~GeV, the remaining background contribution in the signal region is found to be smoothly distributed.

\begin{figure}[!htbp]
    \centering
    \includegraphics[width=0.99\linewidth]{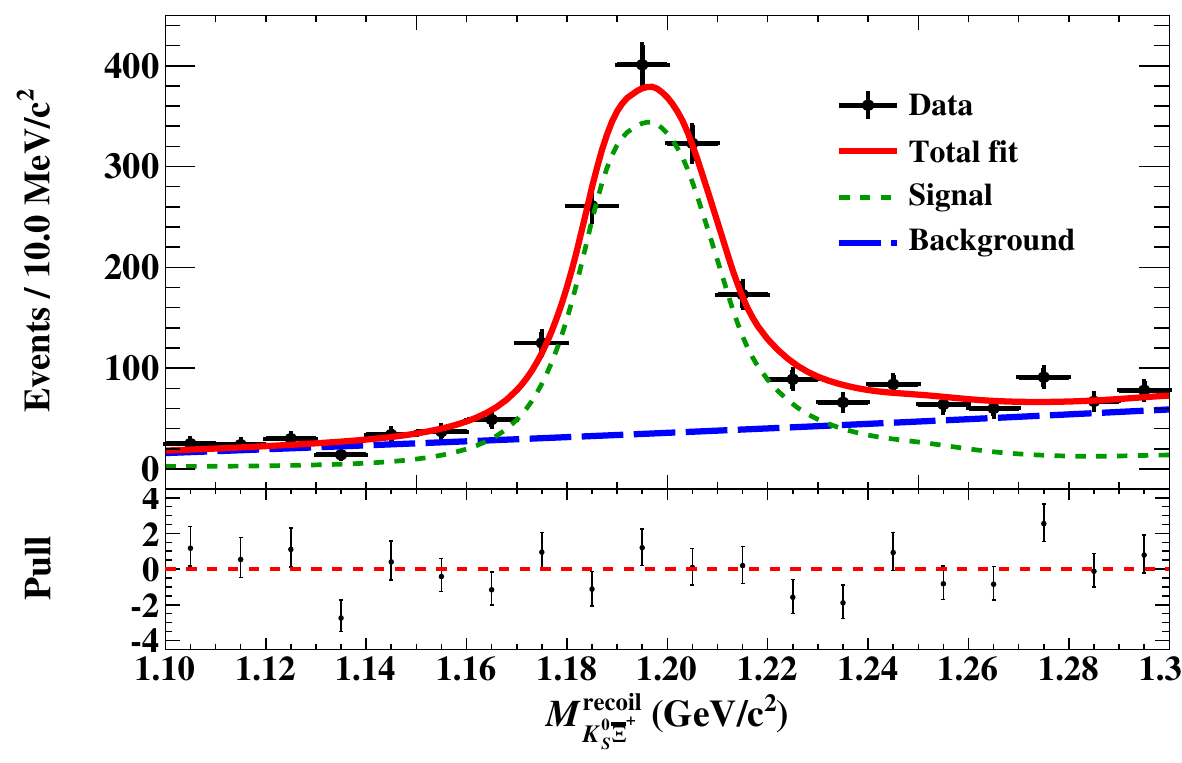}
    \caption{Fit to $M_{K_S^0\bar{\Xi}^+}^{\rm recoil}$ at $\sqrt{s}=$3.773~GeV. Black dots with error bars are data; the red solid line is the total fit; the green short-dashed line is the signal; the blue long-dashed line is the background. The bottom panel shows the pull distribution.}
    \label{fig:fit3773}
\end{figure}

The signal yield for $e^+e^-\to K_S^0\bar{\Xi}^+\Sigma^-$ at each energy point is extracted with an extended maximum-likelihood fit to the $M_{K_S^0\bar{\Xi}^+}^{\rm recoil}$ distribution in the range from 1.1 to 1.3~GeV$/c^2$. 
In the fit, the signal shape is described by the MC-simulated shape convolved with a Gaussian function that accounts for the difference of mass resolution between data and  MC events; for energy points with fewer than 20 events, the pure MC shape is used. Backgrounds are described by first- or second-order polynomials. Figure~\ref{fig:fit3773} demonstrates the fit to the $M_{K_S^0\bar{\Xi}^+}^{\rm recoil}$ distribution at $\sqrt{s} =$ 3.773~GeV. 

The Born cross section ($\sigma^{\rm{B}}$) for $e^+e^-\to K_S^0\bar{\Xi}^+\Sigma^-$ at a given c.m. energy is calculated as
\begin{equation}
    \sigma^{\rm{B}} = \frac{N_{\rm{obs}}}{
    \mathcal{L} \cdot (1+\delta) \cdot \frac{1}{\lvert 1-\Pi \rvert^2} \cdot \varepsilon \cdot \mathcal{B}},
\end{equation}
where $N_{\rm {obs}}$ is the signal yield, $\mathcal{L}$ is the integrated luminosity, $(1+\delta)$ is the ISR correction, $1/|1-\Pi|^2$ is the vacuum-polarization (VP) correction, $\varepsilon$ is the detection efficiency, and $\mathcal{B}$ is the product of the branching fractions for $K_S^0\to \pi^+\pi^-$, $\bar{\Xi}^+\to \bar{\Lambda}\pi^+$, and $\bar{\Lambda} \to \bar{p}\pi^+$ taken from the PDG~\cite{ParticleDataGroup:2022pth}.
The ISR correction is obtained from QED calculations~\cite{Sun:2020ehv}, and the VP correction is computed with a VP computation package~\cite{Jegerlehner:2011ti}.
Efficiencies and ISR factors are determined iteratively as proposed in Ref.~\cite{Sun:2020ehv}.
Numerical results for each energy are collected in Table~\ref{tab:signal:yields:DD} in Appendix.
Figure~\ref{fig:BCS} shows the Born cross section for the reaction of $e^+e^-\to K_S^0\bar{\Xi}^+\Sigma^-$ compared with that of $e^+e^-\to K^-\bar{\Xi}^+\Sigma^0$~\cite{BESIII:2024ogz}, along with their ratio.
\begin{figure}[!htbp]
    \centering
     \includegraphics[width=0.5\textwidth]{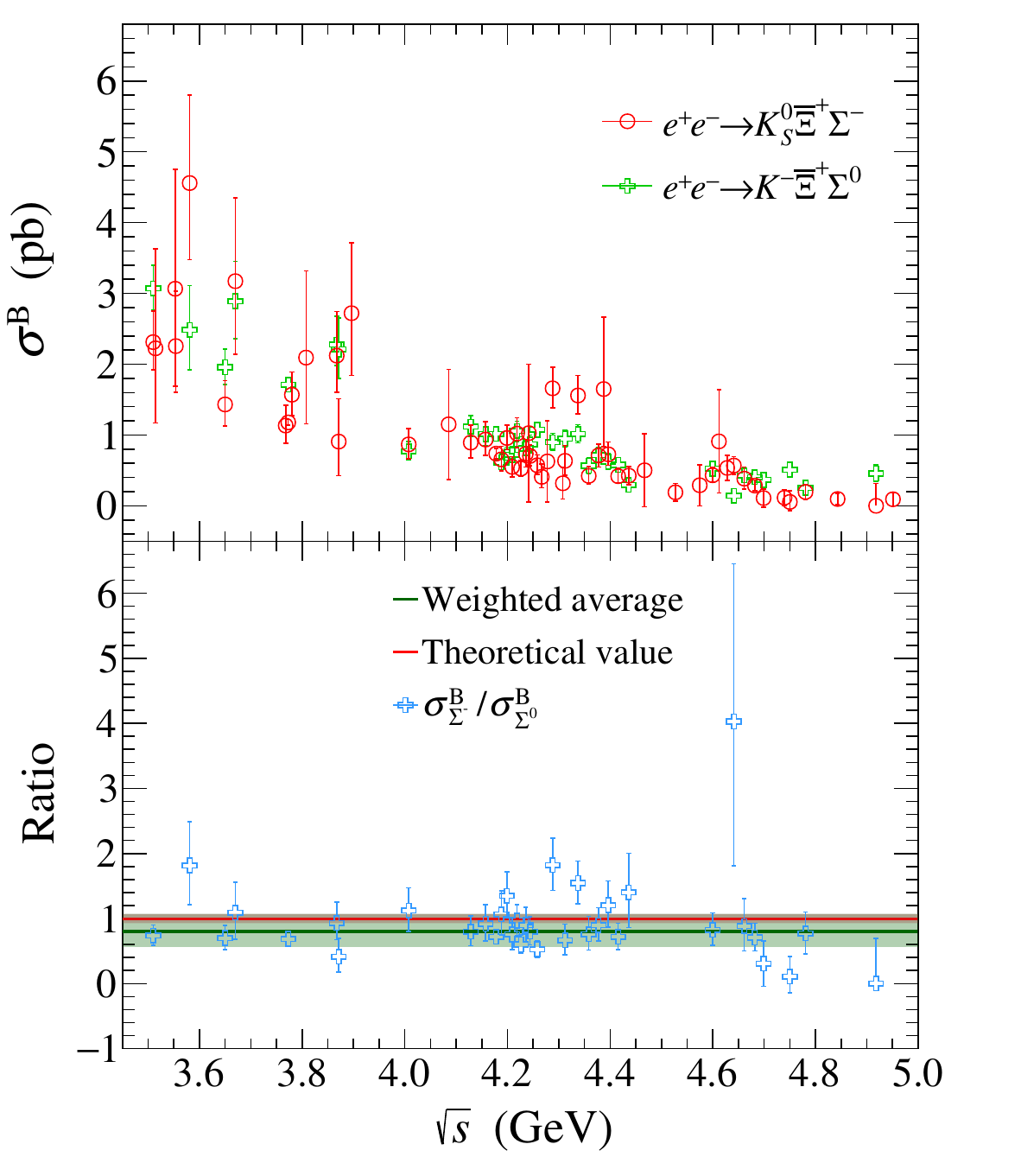}
    \caption{\textbf{Top}: Comparison of the Born cross sections between this work and the previous measurement for the reaction of $e^+e^- \to K^-\bar{\Xi}^+\Sigma^0$~\cite{BESIII:2024ogz} as a function of c.m. energy at $\sqrt{s}$ = 3.510-4.951~GeV. \textbf{Bottom}: Ratio of the Born cross sections for both reactions. The red dashed line with error bars stands for the theoretical calculated value; the solid green line with error bars represents the weighted average value of the measurement.}
    \label{fig:BCS}
\end{figure}

Systematic uncertainties in the cross-section measurement arise from the integrated luminosity $K_S^0$ and $\bar{\Xi}^+$ reconstruction, intermediate-state branching fractions, the physical model, signal and background shapes, and the input lineshape.
The luminosity at all energy points is evaluated using the Bhabha scattering process with an uncertainty of about 1.0\%~\cite{BESIII:2015qfd} below 4.0~GeV, 0.7\%~\cite{BESIII:2022dxl} from 4.0 to 4.6~GeV, and 0.6\%~\cite{BESIII:2022ulv} above 4.6~GeV. 
Reconstruction uncertainties for $K_S^0$ and $\bar{\Xi}^+$ (tracking, PID, mass windows, and decay-length requirements) are evaluated using control samples of $J/\psi\to K^{*\pm}(892)K^{\mp}$, $K^{*\pm}(892)\to K_S^0\pi^\pm$ and $\psi(3686)\to\Xi^{-}\bar\Xi^+$, yielding 2.1\% and 4.4\%, respectively~\cite{BESIII:2021kwf,BESIII:2016ssr,BESIII:2022lsz}.
The uncertainty due to the branching fractions of the intermediate states is taken from the PDG~\cite{ParticleDataGroup:2022pth}. The decays of 
$K_S^0\to\pi^+\pi^-$, $\bar{\Xi}^+\to\bar{\Lambda}\pi^+$ and $\bar{\Lambda} \to \bar{p}\pi^+$, which contribute as 0.8\%.
The model dependence is assessed by comparing efficiencies obtained from the PHSP generator with those from a partial-wave analysis of $e^+e^-\to K_S^0\bar{\Xi}^+\Sigma^-$ at $\sqrt{s}=3.773$~GeV~\cite{BESIII:2022upt,BESIII:2023ojh}, and the 2.5\% difference is taken as the systematic uncertainty.
The uncertainty of the signal shape for the fit of $M_{K_S^0\bar{\Xi}^+}^{\rm recoil}$ is evaluated by combining all energy points and varying the Gaussian parameters within the range of 1$\sigma$, and the difference in the signal yield, 1.3\%, is taken as the systematic uncertainty.
The uncertainty due to the background shape is estimated through an alternative fit with a higher-order polynomial function. The polynomial order is increased from $1^{{\rm{st}}}$ to $2^{\rm{nd}}$ or from $2^{\rm{nd}}$ to $3^{\rm{rd}}$.
The resulting difference of 1.7\% and 0.1\% between the nominal and higher-order fits are taken as the systematic uncertainty.
The uncertainty for input lineshape of cross section has two components: the parameterization and the model. The former is treated as in Refs.~\cite{Sun:2020ehv, BESIII:2024gql}; the latter is estimated by replacing the nominal power-law function with a sum of $\psi(3770)$ Breit-Wigner and power-law functions, recalculating $\varepsilon\cdot(1+\delta)$, and taking the largest deviation (0.6\%).
All individual contributions are added in quadrature to obtain the total systematic uncertainty.

\begin{figure*}[!htbp]
    \centering
    \includegraphics[width=0.5\textwidth]{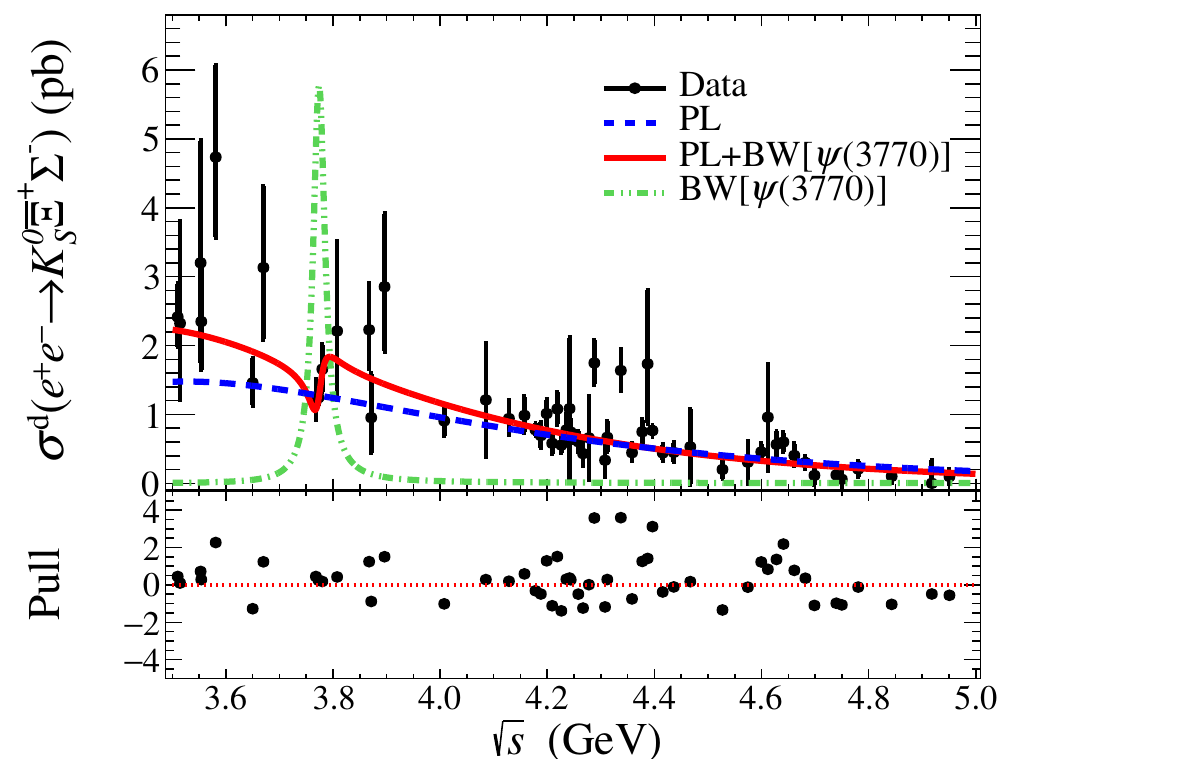}%
    \includegraphics[width=0.5\linewidth]{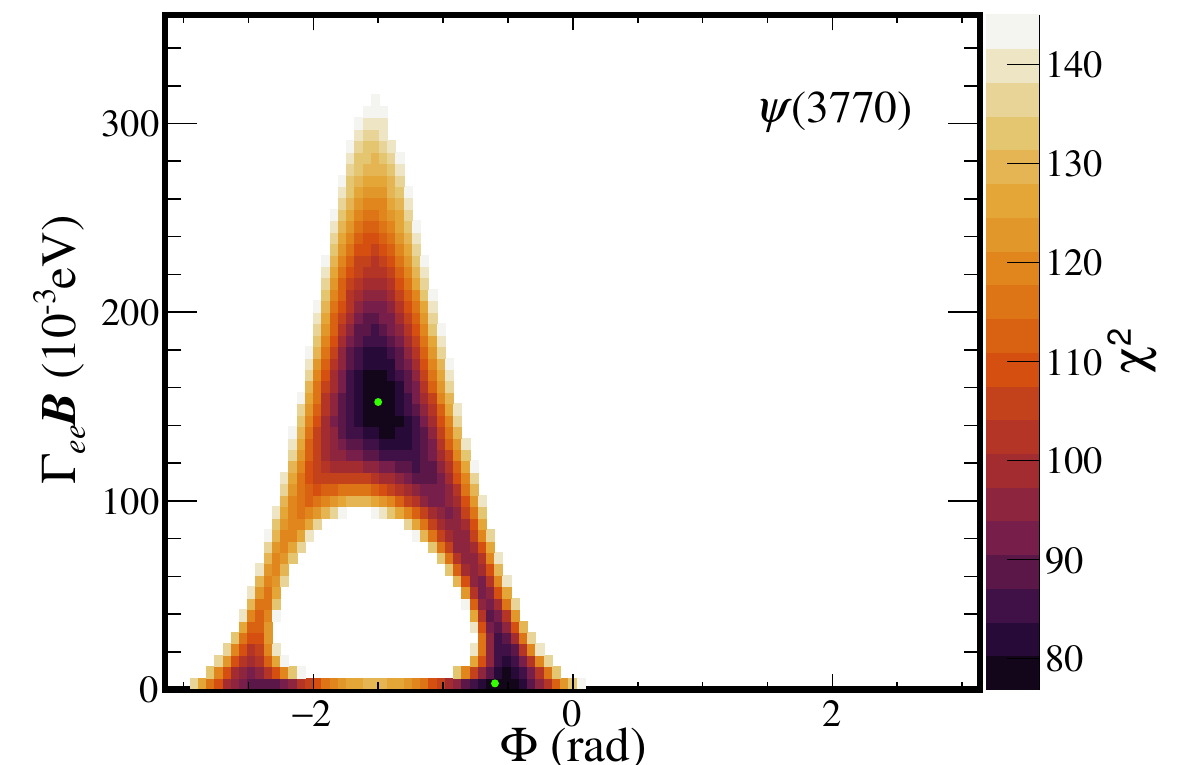}%
    \caption{\textbf{Left}: Fit to the dressed cross sections of $e^+e^-\to K_S^0\bar{\Xi}^+\Sigma^-$. Dots with error bars are the measured dressed cross sections (statistical and systematic uncertainties combined). The blue dashed line is the power-law contribution; the green dashed line is the $\psi(3770)$ signal; the red solid line is the total fit. The bottom panel shows the pull distribution.
    \textbf{Right}: Contours of $\Gamma_{ee}\mathcal{B}$ versus $\Phi$ for $\psi(3770)$. Green markers indicate the best-fit points.}
    \label{fig:psi3770}
\end{figure*}

Resonances or charmonium(-like) states produced in $e^+e^-\to K_S^0\bar{\Xi}^+\Sigma^-$ are searched for by fitting the dressed cross sections $\sigma^{\rm d}=\sigma^{ B}/|1-\Pi|^2$ (including VP) with the least-squares method~\cite{Mo.X.H:2003, Mo:2006bea, Mo:2007aea}.
Systematic uncertainties on luminosity, $K_S^0$, $\bar{\Xi}^+$ reconstruction, branching fractions, MC model and the input lineshape are treated as correlated; those on the fit range, background and signal shapes, and the statistical uncertainties are uncorrelated.
Assuming a sum of resonance as continuum, the dressed cross sections are fitted to the coherent sum of a power-law and a Breit-Wigner (BW) function:
\begin{equation}
    \sigma^{\rm{d}}(\sqrt{s}) = \left \lvert c_0 \frac{\sqrt{P(\sqrt{s})}}{(\sqrt{s})^n}+e^{i\Phi}BW(\sqrt{s})\sqrt{\frac{P(\sqrt{s})}{P(M)}}\right \rvert^2,
    \label{equ::PLBW}
\end{equation}
%where
where $\Phi$ is the relative phase,
\begin{equation}
    BW(\sqrt{s}) = \frac{\sqrt{12\pi\Gamma_{ee}\mathcal{B}\Gamma}}{s-M^2+iM\Gamma},
\end{equation}
$\sqrt{P(\sqrt{s})}$ is the three-body PHSP factor, $M$ and $\Gamma$ are the mass and width of the assumed resonance, i.e. $\psi(3770)$, $\psi(4040)$, $\psi(4160)$, $Y(4230)$, $Y(4360)$, $\psi(4415)$, $Y$(4500)~\cite{BESIII:2022joj}, $Y(4660)$, or $Y$(4710)~\cite{BESIII:2023wqy}, which are fixed to the PDG values~\cite{ParticleDataGroup:2022pth}, the parameters $n$ and $c_0$ are left free in the fit, and $\Gamma_{ee}\mathcal{B}$ is the product of the electronic partial width and the branching fraction for the assumed resonance decaying into the $K_S^0\bar{\Xi}^+\Sigma^-$ final state. 
No significant signal is found, and the upper limits of the $\Gamma_{ee}{\cal{B}}$ for these charmonium(-like) states decaying into the $K_S^0\bar{\Xi}^+\Sigma^-$ final state are provided at the 90\% confidence level (C.L.) using a Bayesian approach~\cite{Zhu:2008ca}. 
Figure~\ref{fig:psi3770} shows the fit to the dressed cross sections and the scan for multiple solutions of the $\psi(3770)$.
The resonance parameters from the fit of different resonances one at a time are summarized in Table~\ref{tab:scan:two}. 
More details can be found in Table~\ref{tab:scan} in Appendix.
Figure~\ref{fig:BCS} shows the Born cross sections for the processes $e^+e^-\to K_S^0\bar{\Xi}^+\Sigma^-$ and $e^+e^-\to K^-\bar{\Xi}^+\Sigma^0$~\cite{BESIII:2024ogz}, as well as their ratio.

Furthermore, the ratio of the Born cross sections is used to test isospin conservation. Based on the assumption of isospin conservation and quark flavor compositions of $\Sigma$ and $\bar{\Xi}$ baryons, the ratio of the Born cross sections for the processes $e^+e^-\to \bar{K}^0\bar{\Xi}^+\Sigma^-$ and $e^+e^-\to K^-\bar{\Xi}^+\Sigma^0$ is predicted to be $R_{\bar{K}^0/K^-}=2$. However, only the mass eigenstate $K_S^0$, associated with the $\pi\pi$ final state, can be reconstructed experimentally.
Given that the branching fraction for $K_S^0\to \pi \pi$ is close to 1~\cite{ParticleDataGroup:2022pth}, we obtain the relation: $K^0_S \approx K_{C\!P=1}$. 
Neutral kaons undergo $\bar{K}^0$-$K^0$ oscillation via the weak interaction, and the $C\!P$ eigenstates of the neutral $K$ mesons are defined as:
\begin{equation}
\begin{aligned}
    K_{C\!P=1}=\frac{1}{\sqrt{2}}(\ket{K^0}+\ket{\bar{K}^0}), \\
    K_{C\!P=-1}=\frac{1}{\sqrt{2}}(\ket{K^0}-\ket{\bar{K}^0}).
\end{aligned}%~\cite{Gell-Mann:1955ipe}. 
\end{equation}
The squared probability amplitude for finding $K_S^0$ in the $\bar{K}^0$ state is given by $P = \bra{K_S^0}\ket{\bar{K}^0}^2 \approx \lvert 1/\sqrt{2}\rvert^2 = 1/2$.
Hence, the ratio of the Born cross sections for $e^+e^-\to K_S^0\bar{\Xi}^+\Sigma^-$ and $e^+e^-\to K^{-}\bar{\Xi}^+\Sigma^0$ is found to be $R_{K_S^0/K^-}=R_{\bar{K}^0/K^-}\cdot~P\approx1$.
The weighted average of the ratio in Fig.~\ref{fig:BCS} is $0.8 \pm 0.2$ , where the uncertainty includes both statistical and systematic contributions. This result is consistent with the isospin conservation assumption.

\begin{table}[!htbp]
    \centering
    \renewcommand\arraystretch{1.18}
    \begin{ruledtabular}
    \caption{Fitted resonance parameters for $\Gamma_{ee}\mathcal{B}$~($10^{-3}$ eV) and $\phi$~(rad) for the two solutions (I, II). Statistical and systematic uncertainties are included. Bracketed values are conservative 90\%~C.L.\ upper limits obtained from $\Delta\chi^2$ scans. The number of degrees of freedom is 51, and $\chi^2/n.d.f$ for the continuum-only fit is 1.7.}
    \scalebox{1.0}{
    \begin{tabular}{cccc} %\hline \hline
    Resonance     & $\Phi$ (rad) & $\Gamma_{ee}\mathcal{B}$ ($\times10^{-3}$ eV) & $\chi^2/n.d.f$ \\ \hline %& $\mathcal{S}(\sigma)$ \\ \hline
    $\psi(3770)^{\rm{I}}$  & $-0.6$$\pm$0.1 & 6.6$^{+4.8}_{-3.2}$ & 1.3 \\ %& 2.1 \\
    $\psi(3770)^{\rm{II}}$  & $-1.4$$\pm$0.1 & 159.6$^{+8.8}_{-8.4}$ ($\textless$173.0) & 1.3 \\ \hline%  & 2.1 \\ \hline
    $\psi(4040)^{\rm{I}}$  & 0.7$\pm$0.9 & 1.2$^{+3.2}_{1.2}$ & 1.5 \\%  & 1.6\\
    $\psi(4040)^{\rm{II}}$  & $-1.6$$\pm$0.1 & 375.3$^{+43.9}_{-44.8}$ ($\textless$473.0) & 1.5 \\ \hline  % & 1.6 \\ \hline
    $\psi(4160)^{\rm{I}}$  & 1.1$\pm$0.9 & 0.0$^{+0.0}_{-0.0}$ & 1.6 \\ %  & 0.6 \\
    $\psi(4160)^{\rm{II}}$  & $-1.6$$\pm$0.3 & 249.6$^{+13.7}_{-12.8}$ ($\textless$275.0)  & 1.6 \\ \hline % & 0.6 \\ \hline
    $Y(4230)^{\rm{I}}$  & $-1.3$$\pm$2.0 & 0.0$^{+0.0}_{-0.0}$ & 1.6 \\ % & 0.2 \\
    $Y(4230)^{\rm{II}}$  & $-1.6$$\pm$0.2 & 164.5$^{+9.9}_{-9.4}$ ($\textless$188.0) & 1.6 \\ \hline %  & 0.2 \\ \hline
    $Y(4360)^{\rm{I}}$  & 1.8$\pm$0.2 & 4.5$^{+2.3}_{-1.8}$ & 1.5 \\  % & 1.6 \\
    $Y(4360)^{\rm{II}}$ & $-1.7$$\pm$0.1 & 372.0$^{+20.7}_{-19.4}$ ($\textless$412.0) & 1.5 \\ \hline % & 1.6 \\ \hline
    $\psi(4415)^{\rm{I}}$ & 2.7$\pm$0.3 & 4.5$^{+2.5}_{-1.8}$ & 1.5 \\ % & 1.5 \\
    $\psi(4415)^{\rm{II}}$  & $-1.8$$\pm$0.1 & 301.5$^{+19.8}_{-19.0}$ ($\textless$350.0) & 1.5 \\ \hline % & 1.5 \\ \hline
    $Y(4500)^{\rm{I}}$ & $-3.1$$\pm$0.7 & 3.6$^{+2.4}_{-1.7}$ & 1.5 \\ %& 2.1 \\
    $Y(4500)^{\rm{II}}$ & $-1.7$$\pm$0.1 & 265.5$^{+29.0}_{-29.7}$ ($\textless$332.0) & 1.5 \\ \hline%  & 2.1 \\ \hline
    $Y(4660)^{\rm{I}}$  & 2.3$\pm$0.3 & 3.7$^{+2.3}_{-1.8}$ & 1.4 \\ % & 2.6 \\
    $Y(4660)^{\rm{II}}$  & $-1.7$$\pm$0.1 & 169.8$^{+12.4}_{-12.1}$ ($\textless$198.0) & 1.4 \\ \hline % & 2.6 \\
    $Y(4710)^{\rm{I}}$ & $-2.4$$\pm$0.2 & 21.3$^{+15.6}_{9.0}$ & 1.5 \\%  & 1.6\\
    $Y(4710)^{\rm{II}}$ & $-1.9$$\pm$0.1 & 157.5$^{+26.3}_{-31.3}$ ($\textless$216.0) & 1.5 \\% \hline
    %\hline \hline
    \end{tabular}
    }
    \label{tab:scan:two}
    \end{ruledtabular}
    
\end{table}

%\section{Summary}
In summary, using a total of 44~fb$^{-1}$ of $e^+e^-$ collision data collected by the BESIII detector at the BEPCII collider, we present the first measurement of the Born cross sections for the $e^+e^-\to K_S^0\bar{\Xi}^+\Sigma^-$ reaction at 56~c.m. energies between 3.510 and 4.951~GeV. 
A fit to the dressed cross sections of $e^+e^-\to K_S^0\bar{\Xi}^+\Sigma^-$ is performed assuming individual charmonium(-like) resonances [$i.e.$, $\psi(3770)$, $\psi(4040)$, $\psi(4160)$, $Y(4230)$, $Y(4360)$, $\psi(4415)$, $Y$(4500), $Y(4660)$, and $Y$(4710)] and a continuum contribution.
No evidence for any possible charmonium(-like) states is found. Upper limits for the products of the branching fraction and the electronic partial width at the 90\% C.L. are provided. 
Based on the Ref.~\cite{BESIII:2024ogz} results, we determine the average ratio of cross sections for two processes to be 
$R=\frac{\sigma^{B}(e^+e^-\to K_S^0\bar{\Xi}^+\Sigma^-)}{\sigma^{B}(e^+e^-\to K^-\bar{\Xi}^+\Sigma^0)}$$ = 0.8\pm0.2$.
This measured ratio is consistent with expectations from isospin conservation and $K^0$-$\bar{K}^0$ oscillation. 
These results offer new experimental insight into three-body baryon production above the open-charm threshold and into the nature of charmonium(-like) states~\cite{Close:2005iz,Wang:2019mhs,Qian:2021neg}.

The BESIII Collaboration thanks the staff of BEPCII (https://cstr.cn/31109.02.BEPC) and the IHEP Computing Center for their strong support. This work is supported in part by National Key R\&D Program of China under Contracts No. 2023YFA1606000, and No. 2023YFA1606704; National Natural Science Foundation of China (NSFC) under Contracts No. 
12247101, No. 11635010, No. 11935015, No. 11935016, No. 11935018, No. 12025502, No. 12035009, No. 12035013, No. 12061131003, No. 12192260, No. 12192261, No. 12192262, No. 12192263, No. 12192264, No. 12192265, No. 12221005, No. 12225509, No. 12235017, No. 12342502, and No. 12361141819; the Fundamental Research Funds for the Central Universities No. lzujbky-2025-ytA05, No. lzujbky-2025-it06, No. lzujbky-2024-jdzx06; the Natural Science Foundation of Gansu Province No. 22JR5RA389, No. 25JRRA799;
the "111 Center" under Grant No. B20063; the Chinese Academy of Sciences (CAS) Large-Scale Scientific Facility Program; the Strategic Priority Research Program of Chinese Academy of Sciences under Contract No. XDA0480600; CAS under Contract No. YSBR-101; 100 Talents Program of CAS; The Institute of Nuclear and Particle Physics (INPAC) and Shanghai Key Laboratory for Particle Physics and Cosmology; ERC under Contract No. 758462; German Research Foundation DFG under Contract No. FOR5327; Istituto Nazionale di Fisica Nucleare, Italy; Knut and Alice Wallenberg Foundation under Contracts No. 2021.0174, No. 2021.0299, No. 2023.0315; Ministry of Development of Turkey under Contract No. DPT2006K-120470; National Research Foundation of Korea under Contract No. NRF-2022R1A2C1092335; National Science and Technology fund of Mongolia; Polish National Science Centre under Contract No. 2024/53/B/ST2/00975; STFC (United Kingdom); Swedish Research Council under Contract No. 2019.04595; U. S. Department of Energy under Contract No. DE-FG02-05ER41374.

\bibliography{apssamp.bib}
\newpage

\newpage
\onecolumngrid

\appendix
%\newpage
\section{End Matter}
\label{app}

This End Matter provides detailed information on the measured Born cross sections for the process $e^+e^-\to K_S^0\bar{\Xi}^+\Sigma^-$.
\begin{table*}[!htbp]
 \centering
    \renewcommand\arraystretch{1.1}
\begin{ruledtabular}
    \caption{The $e^+e^-\to K_S^0\bar{\Xi}^+\Sigma^-$ Born cross sections $\sigma^B$ for 56 energy points between 3.510 and 4.951~GeV. The values in the brackets are the corresponding upper limits at the 90\% C.L.. The first uncertainties are statistical, and the second systematic. The $\sqrt{s}$ is the $e^+e^-$ c.m. energy~\cite{BESIII:2015zbz, BESIII:2020eyu}, $\int\mathcal{L}\rm{d}\it{t}$ is the integrated luminosity of each data set~\cite{BESIII:2015qfd,BESIII:2022dxl, BESIII:2022ulv, Ablikim:2013ntc, BESIII:2024lks, BESIII:2015equ, BESIII:2024lbn}, the vacuum polarization correction factor $\frac{1}{\lvert1-\Pi\rvert^2}$, the detection efficiency $\varepsilon$, and the ISR correction factor ($1+\delta$). $N_{\rm{obs}}$ is the signal yield (the upper limits of signal yields including additive part of systematic uncertainty at the 90\% C.L. at these energy points are determined with the Bayesian method~\cite{Zhu:2008ca}). $\mathcal{S}$ is the statistical significance.}
    \scriptsize
    \centering
    \color{black}{
    \scalebox{0.9}{
   
        \begin{tabular*}{\linewidth}{c@{\extracolsep{\fill}}ccccllc}%\hline \hline
       
		$\sqrt{s}$~(GeV) &$\int\mathcal{L}\rm{d}\it{t}$ (pb$^{-1})$&$\frac{1}{\lvert 1-\Pi \rvert^2}$& (1+$\delta$)& $\varepsilon$(\%)&~~~~$N_{\rm{obs}}$($N^{\rm{UL}}$) &~~~~~$\sigma^{B}$ (fb) & $\mathcal{S}$ ($\sigma$) \\ \hline
    
3.510 &405.7& 1.0444 & 0.7105 & 12.03 &$37.1^{+7.0}_{-6.3}$&$2312^{+436}_{-393}\pm135$& 5.6 \\
3.514 &40.9 & 1.0441 & 0.7331 & 12.32 &$3.8^{+2.4}_{-1.8}~(\textless30.0)$&$2322^{+1475}_{-1111}\pm152~(\textless17561)$& 1.1 \\
3.552 &42.2& 1.0436 & 0.8166 & 12.29 &$6.0^{+3.3}_{-2.7}~(\textless21.8)$&$3200^{+1773}_{-1455}\pm210~(\textless11142)$& 0.9 \\
3.554 &129.4 & 1.0406 & 0.8171 & 13.19 &$14.5^{+5.0}_{-4.2}~(\textless41.1)$&$2347^{+824}_{-697}\pm154~(\textless6370)$& 2.3 \\
3.581 &85.7& 1.0385 & 0.8465 & 12.75 &$19.4^{+5.3}_{-4.6}$&$4559^{+1245}_{-1081}\pm277$& 3.3 \\
3.650 &454.5& 1.0190 & 0.8988 & 13.76 &$37.1^{+8.7}_{-7.9}$&$1432^{+336}_{-305}\pm87$& 4.1 \\
3.670 &84.7& 0.9868 & 0.9105 & 12.62 &$13.5^{+5.0}_{-4.4}~(\textless29.7)$&$3173^{+1175}_{-1034}\pm192~(\textless6980)$& 1.8 \\
3.768 &415.8& 1.0548 & 0.9618 & 13.14 &$27.9^{+7.0}_{-6.2}~(\textless38.9)$&$1136^{+285}_{-252}\pm69~(\textless1580)$& 2.4 \\
3.773 &20249.8& 1.0570 & 0.9624 & 13.42 &$1413.7^{+48.6}_{-47.8}$&$1154^{+40}_{-39}\pm67$& $\gg 5$ \\
3.780 &410& 1.0598 & 0.9657 & 13.24 &$38.5^{+7.9}_{-7.2}$&$1563^{+321}_{-292}\pm91$& 4.4 \\
3.808 &50.5& 1.0561 & 0.9778 & 13.43 &$6.5^{+3.8}_{-2.9}~(\textless21.1)$&$2093^{+1224}_{-934}\pm127~(\textless6795)$& 1.9 \\
3.867 &108.9& 1.0507 & 1.0032 & 12.98 &$14.0^{+4.1}_{-3.4}$&$2121^{+621}_{-515}\pm129$& 4.0 \\
3.871 &110.3& 1.0504 & 1.0053 & 12.81 &$6.0^{+4.0}_{-3.2}~(\textless22.0)$&$907^{+605}_{-484}\pm55~(\textless3327)$& 1.5 \\
3.896 &52.6& 1.0486 & 1.0154 & 13.33 &$9.0^{+3.3}_{-2.9}~(\textless23.8)$&$2720^{+997}_{-877}\pm165~(\textless7194)$& 2.7 \\
4.008 &482& 1.0438 & 1.0651 & 13.21 &$27.2^{+6.9}_{-6.3}$&$867^{+220}_{-201}\pm52$& 3.8 \\
4.085 &52.9& 1.0515 & 1.0966 & 12.86 &$4.0^{+2.7}_{-2.7}~(\textless8.4)$&$1151^{+777}_{-777}\pm67~(\textless2416)$& 1.8 \\
4.128 &401.5& 1.0526 & 1.1178 & 11.77 &$22.0^{+6.0}_{-5.3}$&$893^{+244}_{-215}\pm54$& 3.5 \\
4.157 &408.7& 1.0535 & 1.1332 & 11.77 &$23.8^{+6.4}_{-5.7}~(\textless47.8)$&$936^{+252}_{-224}\pm56~(\textless1879)$& 2.9 \\
4.178 &3189& 1.0543 & 1.1415 & 12.11 &$151.7^{+15.5}_{-14.7}$&$737^{+75}_{-71}\pm43$& 8.7 \\
4.188 &526.7& 1.0559 & 1.1465 & 12.20 &$22.6^{+6.3}_{-5.5}$&$656^{+183}_{-160}\pm40$& 3.0 \\
4.199 &526& 1.0567 & 1.1516 & 12.37 &$33.6^{+6.4}_{-5.8}$&$958^{+182}_{-165}\pm58$& 4.2 \\
4.209 &517.1& 1.0565 & 1.1576 & 11.95 &$18.4^{+4.8}_{-4.8}~(\textless35.8)$&$550^{+143}_{-143}\pm33~(\textless1069)$& 2.3 \\
4.219 &514.6& 1.0564 & 1.1609 & 11.91 &$34.0^{+7.4}_{-6.7}$&$1021^{+222}_{-201}\pm59$& 4.1 \\
4.226 &1100.9& 1.0562 & 1.1643 & 12.69 &$40.0^{+8.6}_{-7.9}$&$525^{+113}_{-104}\pm32$& 3.7 \\
4.236 &530.3& 1.0554 & 1.1703 & 12.35 &$26.3^{+7.3}_{-6.6}$&$734^{+204}_{-184}\pm44$& 3.0 \\
4.242 &55.88& 1.0555 & 1.1742 & 12.70 &$4.0^{+3.8}_{-3.8}~(\textless8.4)$&$1026^{+975}_{-975}\pm62~(\textless2156)$& 2.2 \\
4.244 &538.1& 1.0555 & 1.1743 & 12.20 &$25.4^{+5.7}_{-5.0}$&$704^{+158}_{-139}\pm41$& 3.3 \\
4.258 &828.4& 1.0535 & 1.1818 & 12.49 &$32.7^{+8.2}_{-7.5}~(\textless55.3)$&$573^{+144}_{-131}\pm33~(\textless969)$& 2.8 \\
4.267 &531.1& 1.0532 & 1.1867 & 12.13 &$14.6^{+6.4}_{-5.6}~(\textless30.7)$&$409^{+179}_{-157}\pm25~(\textless861)$& 1.2 \\
4.278 &175.7& 1.0530 & 1.1920 & 11.76 &$7.2^{+6.6}_{-6.6}~(\textless12.5)$&$627^{+574}_{-574}\pm38~(\textless1088)$& 2.0 \\
4.288 &502.4& 1.0526 & 1.1981 & 11.08 &$51.6^{+9.3}_{-8.5}$&$1659^{+299}_{-273}\pm100$& 4.5 \\
4.308 &45.08& 1.0523 & 1.2066 & 12.40 &$1.0^{+0.7}_{-0.7}~(\textless3.7)$&$318^{+223}_{-223}\pm19~(\textless1177)$& 0.8 \\
4.312 &501.2& 1.0521 & 1.2102 & 11.14 &$20.0^{+6.9}_{-6.3}~(\textless38.8)$&$635^{+219}_{-200}\pm38~(\textless1232)$& 1.7 \\
4.337 &505.0& 1.0508 & 1.2238 & 11.26 &$50.5^{+9.1}_{-8.5}$&$1558^{+281}_{-262}\pm94$& 4.7 \\
4.358 &543.9& 1.0511 & 1.2335 & 12.27 &$16.2^{+5.0}_{-4.3}~(\textless22.3)$&$423^{+130}_{-112}\pm24~(\textless582)$& 2.3 \\
4.377 &522.7& 1.0514 & 1.2459 & 11.08 &$23.9^{+5.4}_{-5.4}$&$711^{+161}_{-161}\pm41$& 3.2 \\
4.387 &55.6& 1.0513 & 1.2517 & 12.16 &$6.5^{+4.0}_{-3.2}~(\textless9.4)$&$1649^{+1015}_{-812}\pm99~(\textless2385)$& 1.6 \\
4.396 &507.8& 1.0513 & 1.2582 & 11.00 &$23.7^{+5.7}_{-5.1}$&$724^{+174}_{-156}\pm42$& 3.6 \\
4.416 &1090.7& 1.0524 & 1.2684 & 11.90 &$32.1^{+8.4}_{-7.6}~(\textless49.2)$&$418^{+109}_{-99}\pm25~(\textless641)$& 2.7 \\
4.436 &569.9& 1.0541 & 1.2803 & 10.89 &$15.8^{+4.8}_{-4.8}~(\textless32.9)$&$426^{+129}_{-129}\pm25~(\textless887)$& 2.7 \\
4.467 &111.1& 1.0548 & 1.3008 & 11.51 &$3.9^{+4.0}_{-4.0}~(\textless6.1)$&$502^{+515}_{-515}\pm30~(\textless785)$& 1.3 \\
4.527 &112.1& 1.0545 & 1.3402 & 11.23 &$1.5^{+1.0}_{-1.0}~(\textless3.7)$&$190^{+127}_{-127}\pm11~(\textless469)$& 0.3 \\
4.575 &48.9& 1.0545 & 1.3762 & 10.94 &$1.0^{+1.0}_{-1.0}~(\textless3.7)$&$290^{+290}_{-290}\pm17~(\textless1075)$& 1.1 \\
4.600 &586.9& 1.0547 & 1.3928 & 10.93 &$18.0^{+4.6}_{-4.0}$&$431^{+110}_{-96}\pm26$& 3.4 \\
4.612 &103.7& 1.0545 & 1.4011 & 9.72 &$6.0^{+4.8}_{-4.8}~(\textless11.2)$&$910^{+728}_{-728}\pm55~(\textless1698)$& 2.2 \\
4.628 &521.5& 1.0544 & 1.4125 & 9.39 &$17.3^{+5.7}_{-5.7}~(\textless34.5)$&$535^{+176}_{-176}\pm32~(\textless1067)$& 2.6 \\
4.641 &551.7& 1.0544 & 1.4231 & 9.55 &$19.9^{+4.4}_{-4.4}$&$568^{+126}_{-126}\pm34$& 3.3 \\
4.661 &529.4& 1.0544 & 1.4424 & 9.35 &$12.8^{+5.4}_{-4.8}~(\textless29.1)$&$384^{+162}_{-144}\pm23~(\textless872)$& 1.6 \\
4.682 &1667.4& 1.0545 & 1.4558 & 9.31 &$30.2^{+8.5}_{-8.0}~(\textless31.7)$&$286^{+80}_{-76}\pm17~(\textless300)$& 2.1 \\
4.699 &535.5& 1.0545 & 1.4694 & 9.21 &$3.7^{+4.3}_{-4.3}~(\textless6.1)$&$109^{+127}_{-127}\pm6~(\textless180)$& 0.5 \\
4.740 &163.9& 1.0548 & 1.5008 & 9.85 &$1.3^{+1.3}_{-1.3}~(\textless4.2)$&$115^{+115}_{-115}\pm7~(\textless371)$& 0.6 \\
4.750 &366.6& 1.0549 & 1.5098 & 9.87 &$1.4^{+4.0}_{-3.2}~(\textless29.8)$&$55^{+157}_{-125}\pm3~(\textless1168)$& 0.3 \\
4.781 &511.5& 1.0552 & 1.5365 & 9.74 &$7.0^{+3.9}_{-3.9}~(\textless14.9)$&$196^{+109}_{-109}\pm12~(\textless416)$& 1.8 \\
4.843 &525.16& 1.0557 & 1.6012 & 9.27 &$3.5^{+3.0}_{-3.0}~(\textless6.1)$&$96^{+82}_{-82}\pm6~(\textless167)$& 0.9 \\
4.918 &207.82& 1.0562 & 1.6742 & 8.88 &$0.0^{+4.6}_{-0.0}~(\textless7.7)$&$0^{+318}_{-0}\pm0~(\textless533)$& 1.0 \\
4.951 &159.3& 1.0564 & 1.7086 & 8.47 &$1.0^{+1.0}_{-1.0}~(\textless9.0)$&$93^{+93}_{-93}\pm5~(\textless834)$& 0.9 \\

    \end{tabular*}
    
    }
    }
    \label{tab:signal:yields:DD}
    \end{ruledtabular}

\end{table*}

\begin{table}[!htp]
    \centering
    \renewcommand\arraystretch{1.1}
    %\color{red}
    \caption{The fitted resonance parameters for the decay of $e^+e^-\to K_S^0\bar{\Xi}^+\Sigma^-$ with two solutions (I, II), including systematic uncertainty except for center-of-mass energy calibration.}
    
    \scalebox{1.0}{
    \begin{tabular}{ccccccccc} \hline \hline
    Resonance     & $M$(GeV) & $\Gamma$(GeV) & $\Phi$(rad) & $\Gamma_{ee}\mathcal{B}$($\times10^{-3}$eV) & $\mathcal{B}$($\times10^{-6}$) & $n$ & $c_0$  & $\mathcal{S}(\sigma)$ \\ \hline
    $\psi(3770)^{\rm{I}}$ & 3.7737 (fixed) & 0.0272 (fixed) & -0.6$\pm$0.1 & 6.6$^{+4.8}_{-3.2}$ & 25.3$^{+18.4}_{-12.3}$ & 7.9$^{+0.2}_{-0.2}$ & 199.6$^{+61.9}_{-48.0}$  & 0.9 \\
    $\psi(3770)^{\rm{II}}$ & 3.7737 (fixed) & 0.0272 (fixed) & -1.4$\pm$0.1 & 159.6$^{+8.8}_{-8.4}$ ($\textless$173.0) & 611.2$^{+33.7}_{-32.2}$ ($\textless$662.5) & 7.9$^{+0.2}_{-0.2}$ & 202.9$^{+62.9}_{-48.9}$  & 0.9 \\ %\hline
    $\psi(4040)^{\rm{I}}$ & 4.0400 (fixed) & 0.0840 (fixed) & 0.7$\pm$0.9 & 1.2$^{+3.2}_{-1.2}$ & 1.4$^{+3.7}_{-1.4}$ & 6.9$^{+2.2}_{-2.1}$ & 46.0$^{+15.8}_{-11.4}$  & 0.3\\
    $\psi(4040)^{\rm{II}}$ & 4.0400 (fixed) & 0.0840 (fixed) & -1.6$\pm$0.1 & 375.3$^{+43.9}_{-44.8}$ ($\textless$473.0) & 438.4$^{+51.3}_{-52.3}$ ($\textless$552.6) & 6.9$^{+2.2}_{-2.1}$ & 48.8$^{+17.4}_{-12.2}$  & 0.3 \\% \hline
    $\psi(4160)^{\rm{I}}$ & 4.1910 (fixed) & 0.0690 (fixed) & 1.1$\pm$0.9 & 0.0$^{+0.0}_{-0.0}$ & 0.0$^{+0.0}_{-0.0}$ & 6.7$^{+1.6}_{-1.5}$ & 36.4$^{+9.0}_{-6.4}$  & 0.0 \\
    $\psi(4160)^{\rm{II}}$ & 4.1910 (fixed) & 0.0690 (fixed) & -1.6$\pm$0.3 & 249.6$^{+13.7}_{-12.8}$ ($\textless$275.0) & 524.3$^{+28.8}_{-26.9}$ ($\textless$577.6) & 6.7$^{+1.6}_{-1.6}$ & 37.2$^{+9.3}_{-7.2}$  & 0.0 \\ %\hline
    $Y(4230)^{\rm{I}}$ & 4.2222 (fixed) & 0.0510 (fixed) & -1.3$\pm$2.0 & 0.0$^{+0.0}_{-0.0}$ & $-$ & 6.7$^{+0.1}_{-0.1}$ & 34.5$^{+7.3}_{-5.9}$  & 0.0 \\
    $Y(4230)^{\rm{II}}$ & 4.2222 (fixed) & 0.0510 (fixed) & -1.6$\pm$0.2 & 164.5$^{+9.9}_{-9.4}$ ($\textless$188.0) & $-$ & 6.7$^{+0.1}_{-0.1}$ & 34.7$^{+7.3}_{-6.0}$  & 0.0 \\ %\hline
    $Y(4360)^{\rm{I}}$ & 4.3740 (fixed) & 0.1200 (fixed) & 1.8$\pm$0.2 & 4.5$^{+2.3}_{-1.8}$ & $-$ & 7.1$^{+0.2}_{-0.2}$ & 62.0$^{+17.7}_{-13.3}$ & 2.8 \\
    $Y(4360)^{\rm{II}}$ & 4.3740 (fixed) & 0.1200 (fixed) & -1.7$\pm$0.1 & 372.0$^{+20.7}_{-19.4}$ ($\textless$412.0) & $-$ & 7.1$^{0.2}_{-0.2}$ & 65.7$^{+19.4}_{-14.4}$  & 2.8 \\ %\hline
    $\psi(4415)^{\rm{I}}$ & 4.4150 (fixed) & 0.1100 (fixed) & 2.7$\pm$0.3 & 4.5$^{+2.5}_{-1.8}$ & 4.3$^{+2.4}_{-1.8}$ & 7.0$^{+0.2}_{-0.2}$ & 50.3$^{+13.7}_{-10.4}$  & 2.9 \\
    $\psi(4415)^{\rm{II}}$ & 4.4150 (fixed) & 0.1100 (fixed) & -1.8$\pm$0.1 & 301.5$^{+19.8}_{-19.0}$ ($\textless$350.0) & 291.6$^{+19.1}_{-18.4}$ ($\textless$338.5) & 7.0$^{+0.2}_{-0.2}$ & 52.3$^{+14.4}_{-11.0}$  & 2.9 \\% \hline
    $Y(4500)^{\rm{I}}$ & 4.4847 (fixed) & 0.1111 (fixed) & $-3.1$$\pm$0.7 & 3.6$^{+2.4}_{-1.7}$ & $-$ & $6.7^{+0.1}_{-0.1}$ & $36.4^{+7.6}_{-6.2}$ & 2.0 \\ %& 2.1 \\
    $Y(4500)^{\rm{II}}$ & 4.4847 (fixed) & 0.1111 (fixed) & $-1.7$$\pm$0.1 & 265.5$^{+29.0}_{-29.7}$ ($\textless$332.0) & $-$ & $6.7^{+0.1}_{-0.1}$ & $36.4^{+7.6}_{-6.2}$ & 2.0 \\ %\hline%  & 2.1 \\ \hline
    $Y(4660)^{\rm{I}}$ & 4.6230 (fixed) & 0.0550 (fixed) & 2.3$\pm$0.3 & 3.7$^{+2.3}_{-1.8}$ & $-$ & 6.8$^{+1.4}_{-1.4}$ & 42.6$^{+9.3}_{-7.5}$  & 2.7 \\
    $Y(4660)^{\rm{II}}$ & 4.6230 (fixed) & 0.0550 (fixed) & -1.7$\pm$0.1 & 169.8$^{+12.4}_{-12.1}$ ($\textless$198.0) & $-$ & 6.9$^{+0.1}_{-0.1}$ & 42.8$^{+9.4}_{-7.6}$  & 2.7 \\ %\hline
    $Y(4710)^{\rm{I}}$ & 4.7080 (fixed) & 0.1260 (fixed) & $-2.4$$\pm$0.2 & 21.3$^{+15.6}_{-9.0}$ & $-$ & $6.6^{+0.2}_{-0.2}$ & $32.3^{+8.4}_{-6.5}$ & 0.8 \\%  & 1.6\\
    $Y(4710)^{\rm{II}}$ & 4.7080 (fixed) & 0.1260 (fixed) & $-1.9$$\pm$0.1 & 157.5$^{+26.3}_{-31.3}$ ($\textless$216.0) & $-$ & $6.6^{+0.2}_{-0.2}$ & $32.7^{+8.5}_{-6.7}$ & 0.8 \\   % & 1.6 \\ \hline
    \hline \hline
    \end{tabular}
    
    }
    \label{tab:scan}
\end{table}

\onecolumngrid
\newpage
\begin{center}
    \newpage
%% Saved at => 2025-11-27
M.~Ablikim$^{1}$\BESIIIorcid{0000-0002-3935-619X},
M.~N.~Achasov$^{4,c}$\BESIIIorcid{0000-0002-9400-8622},
P.~Adlarson$^{82}$\BESIIIorcid{0000-0001-6280-3851},
X.~C.~Ai$^{88}$\BESIIIorcid{0000-0003-3856-2415},
C.~S.~Akondi$^{31A,31B}$\BESIIIorcid{0000-0001-6303-5217},
R.~Aliberti$^{39}$\BESIIIorcid{0000-0003-3500-4012},
A.~Amoroso$^{81A,81C}$\BESIIIorcid{0000-0002-3095-8610},
Q.~An$^{78,64,\dagger}$,
Y.~H.~An$^{88}$\BESIIIorcid{0009-0008-3419-0849},
Y.~Bai$^{62}$\BESIIIorcid{0000-0001-6593-5665}
% ...
\\
\vspace{0.2cm}
(BESIII Collaboration)\\
\vspace{0.2cm} {\it
$^{1}$ Institute of High Energy Physics, Beijing 100049, People's Republic of China\\
% ...
$^{4}$ Budker Institute of Nuclear Physics SB RAS (BINP), Novosibirsk 630090, Russia\\
% ...
$^{31}$ INFN Sezione di Ferrara, (A)INFN Sezione di Ferrara, I-44122, Ferrara, Italy; (B)University of Ferrara, I-44122, Ferrara, Italy\\
% ...
$^{39}$ Johannes Gutenberg University of Mainz, Johann-Joachim-Becher-Weg 45, D-55099 Mainz, Germany\\
% ...
$^{62}$ Southeast University, Nanjing 211100, People's Republic of China\\
% ...
$^{64}$ State Key Laboratory of Particle Detection and Electronics, Beijing 100049, Hefei 230026, People's Republic of China\\
% ...
$^{78}$ University of Science and Technology of China, Hefei 230026, People's Republic of China\\
% ...
$^{81}$ University of Turin and INFN, (A)University of Turin, I-10125, Turin, Italy; (B)University of Eastern Piedmont, I-15121, Alessandria, Italy; (C)INFN, I-10125, Turin, Italy\\
$^{82}$ Uppsala University, Box 516, SE-75120 Uppsala, Sweden\\
% ...
$^{88}$ Zhengzhou University, Zhengzhou 450001, People's Republic of China\\

\vspace{0.2cm}
$^{\dagger}$ Deceased\\
% ...
$^{c}$ Also at the Novosibirsk State University, Novosibirsk, 630090, Russia\\
% ...

}
%% ends here %%

\end{center}

\end{document}